\documentstyle[sprocl,epsfig]{article}

\def\Journal#1#2#3#4{{#1} {\bf #2}, #3 (#4)}

\def\NP{\em Nucl.\ Phys.} 
\def\PL{\em Phys.\ Lett.} 
\def\PP{\em Phys.\ Rep.}
\def\PR{\em Phys.\ Rev.\ Lett.}
\def\PR{\em Phys.\ Rev.} 
\def\ZP{\em Z.\ Phys.} 
\def\MP{\em Int.\ J. Mod.\ Phys.} 
\def\JP{\em J.\ Phys.} 
\newcommand{\nn}{\nonumber}

\newcommand{\beq}{\begin{equation}}
\newcommand{\eeq}{\end{equation}}
\newcommand{\bea}{\begin{eqnarray}}
\newcommand{\eea}{\end{eqnarray}}
\newcommand{\ra}{\rightarrow}
\newcommand{\gsim}{\raisebox{-0.04cm}{$\:\stackrel{>}{{\scriptstyle
 \sim}}\: $} }
\newcommand{\lsim}{\raisebox{-0.04cm}{$\:\stackrel{<}{{\scriptstyle
 \sim}}\: $} }
\newcommand{\MSb}{$\overline{\mbox{MS}}$}
\newcommand{\DISg}{$\mbox{DIS}_{\gamma}$}

\begin{document}

\begin{titlepage}

\begin{flushleft}
WUE-ITP-97-036 \hfill 
August 1997 \\
{\tt hep-ph/9709345} 
\end{flushleft}

\setcounter{page}{0}

\vspace*{\fill}
\begin{center}
{\Large\bf The Parton Structure of Real Photons$^{\,\ast}\!\!\!$}\\

\vspace{5em}
\large
Andreas Vogt

\vspace{2em}
\normalsize
{\it Institut f\"ur Theoretische Physik, Universit\"at W\"urzburg}\\
\normalsize{\it Am Hubland, D--97074 W\"urzburg, Germany} \\

\vspace*{3em}
\vspace*{\fill}
 
{\large \bf Abstract}
\end{center}
\vspace*{3mm}
The QCD treatment of the photon structure is recalled. Emphasis is given to 
the recently derived momentum sum rule, and to the proper choice of the
factorization scheme and/or boundary conditions for the evolution equations
beyond the leading order. Parametrizations of the photon's parton content are 
examined and compared. The small-$x$ behaviour of the photon structure is 
briefly discussed.

\noindent

\vspace*{\fill}
\noindent
$^{\ast}$ Invited talk presented at {\sc Photon '97}, Egmond aan Zee, 
The Netherlands, May 1997. To appear in the proceedings.

\newpage
\thispagestyle{empty}
\setcounter{page}{0}
$\,$

\end{titlepage}

\title{THE PARTON STRUCTURE OF REAL PHOTONS}

\author{ANDREAS VOGT}

\address{Institut f\"ur Theoretische Physik, Universit\"at W\"urzburg, \\
         Am Hubland, D--97074 W\"urzburg, Germany}

\maketitle\abstracts{
The QCD treatment of the photon structure is recalled. Emphasis is given to 
the recently derived momentum sum rule, and to the proper choice of the
factorization scheme and/or the boundary conditions for the evolution equations
beyond the leading order. Parametrizations of the photon's parton content are 
examined and compared. The small-$x$ behaviour of the photon structure is 
briefly discussed.}

\section{Introduction}
Deep--inelastic electron--photon scattering has been the classical process 
for investigating the hadronic structure of the photon~\cite{Revs}. This 
process is kinematically analogous to the usual lepton--nucleon scattering. It 
has quite early received special interest, since the structure function 
$F_2^{\,\gamma}(x,Q^2)$ can be completely calculated in perturbation 
theory~\cite{Wit,BB} at large Bjorken-$x$ and large resolution $Q^2$. At scales 
accessible at present or in the near future, however, these results are 
unfortunately not applicable. Hence the photon structure functions have to be 
analyzed in terms of non--perturbative initial distributions for the QCD 
evolution equations \cite{GGR}, very much like the nucleon case.

Experimentally $F_2^{\,\gamma}$ has been determined, albeit with rather limited
accuracy, via $e^+e^- \ra e^+e^- + $ {\it hadrons} at all electron--positron
colliders since PEP and PETRA. The longitudinal structure function $F_L^{\,
\gamma}$ has been unaccessible so far, and will presumably remain so in the 
foreseeable future~\cite{LEP2,Mi97}. On the other hand, the past months have 
witnessed a substantial amount of new results on $F_{2}^{\,\gamma}$ from LEP,
and many more can be expected from forthcoming LEP2 runs. If systematic 
problems in extractions of $F_{2}^{\,\gamma}$ from final--state modeling~\cite
{Mi97} can be overcome, these results will be able to challenge seriously the 
present, model--driven theoretical understanding of the photon structure.

In this talk a brief survey is given of the present theoretical and 
phenomenological status of this subject. In Section 2 we recall the evolution 
equations for the photon's parton distributions, including the recently derived
momentum sum rule. The factorization scheme ambiguities are more relevant here 
as in the usual hadronic case, this issue is discussed in Section 3. Some of 
the most relevant parametrizations of the quark and gluon densities of the 
photon are discussed in Section 4 with respect to their assumptions and 
limitations. Finally Section 5 is devoted to the small-$x$ behaviour of the
photon structure functions. For other aspects the reader is referred to
refs.~\cite{Revs}.
\section{The evolution of the photon's parton densities}
The photon is a genuine elementary particle, unlike the hadrons. Hence it can
directly take part in hard scattering processes, in addition to its quark and
gluon distributions arising from quantum fluctuations, $q^{\gamma}(x,Q^2)$ and
$g^{\gamma}(x,Q^2)$. Denoting the corresponding photon distribution in the 
photon by $\Gamma^{\,\gamma}(x,Q^2)$, the evolution equations for these parton 
densities are generally given by
\bea
  \frac{dq_i^{\,\gamma}}{d\ln Q^{2}} &\! =\! & 
    \frac{\alpha}{2\pi} \overline{P}_{q_{i}\gamma} \otimes \Gamma^{\,\gamma}
    \, + \,\frac{\alpha_{s}}{2\pi} \bigg\{ 2 \sum_{k=1}^{f} 
      \overline{P}_{q_{i}q_{k}} \otimes q_{k}^{\,\gamma}
    \, + \,\overline{P}_{q_{i}g} \otimes g^{\gamma} \bigg\}     \nn \\
\label{evol1}
  \frac{dg^{\gamma}}{d\ln Q^{2}} &\! =\! &
    \frac{\alpha}{2\pi} \overline{P}_{g\gamma} \:\otimes \Gamma^{\,\gamma}
    \, + \,\frac{\alpha_{s}}{2\pi} \bigg\{ 2 \sum_{k=1}^{f} 
      \overline{P}_{gq_{k}} \:\otimes q_{k}^{\,\gamma}
    \, + \,\overline{P}_{gg} \:\otimes g^{\gamma} \bigg\}       \\  
  \frac{d\,\Gamma^{\,\gamma}}{d\ln Q^{2}} &\! =\! &
    \frac{\alpha}{2\pi} \overline{P}_{\gamma\gamma} \:\otimes \Gamma^{\,\gamma}
    \, + \,\frac{\alpha}{2\pi} \bigg\{ 2 \sum_{k=1}^{f}
      \overline{P}_{\gamma q_{k}} \:\otimes q_{k}^{\,\gamma}
    \, + \,\overline{P}_{\gamma g} \:\otimes g^{\gamma} \bigg\} \: . \nn 
\eea
Here $\alpha \simeq 1/137$ is the electromagnetic coupling constant, and 
$\alpha_s \equiv \alpha_s(Q^2)$ denotes the running QCD coupling. $\otimes $ 
represents the Mellin convolution, and $f$ stands for the number of active 
(massless) quark flavours. The antiquark distributions do not occur separately 
in Eq.~(\ref{evol1}), as $\bar{q}_i ^{\,\gamma}(x,Q^2) = q_i^{\,\gamma}(x,Q^2)$ 
due to charge conjugation invariance. The generalized splitting functions read
\beq
\label{pgen}
  \overline{P}_{ij}(x,\alpha, \alpha_s) = \sum_{l,m=0} \frac{ \alpha^l 
  \alpha_s^m}{(2\pi )^{l+m} } \overline{P}_{ij}^{\, (l,m)}(x) \: ,
\eeq
with $\overline{P}_{q_{i}q_{k}}$ being the average of the quark--quark and
antiquark--quark splitting functions. The parton densities are subject to the 
energy--momentum sum rule
\beq
\label{msr1}
  \int_0^1 \! dx\: x \Big[ \Sigma^{\,\gamma}(x,Q^2) + g^{\gamma}(x,Q^2) + 
  \Gamma^{\,\gamma}(x,Q^2) \Big] = 1 \: ,
\eeq
where $\Sigma $ represents the singlet quark distribution, $\Sigma^{\,\gamma}
 = 2 \sum_{i=1}^{f} q_i^{\,\gamma}$. 

Usually calculations involving the photon's parton structure are restricted
to first order in $\alpha \ll 1$. In this approximation all $l\neq 0$ terms in 
Eq.~(\ref{pgen}) can be neglected, since $ q_i^{\,\gamma}$ and $g^{\gamma}$ are 
already of order $\alpha$. This reduces the functions $\overline {P}_{ij}$ to 
the usual QCD quantities ${P}_{ij}(x,\alpha_s)$, with ${P}_{\gamma q_i}$ and 
${P}_{\gamma g}$ dropping out completely. Moreover one has ${P}_{\gamma \gamma} 
\propto \delta (1-x)$ to all orders in $\alpha_s$, as real photon radiations 
from photons starts at order $\alpha^2$ only. Thus the last line of Eq.~(\ref
{evol1}) can be integrated immediately, at leading order (LO), $m=0$, resulting 
in 
\beq
\label{evol2}
  \Gamma_{\rm LO}^{\,\gamma}(x,Q^2) = \delta (1-x) \Big[ 1 - \frac{\alpha}{\pi} 
  \Big( \sum_q e_q^2 \,\ln \frac{Q^2}{Q_0^2} + c_1 \Big) \Big] \: .
\eeq
Here $e_q$ stands for the quark charges, $Q_0^2$ is some reference scale for 
the evolution, and the constant $c_1$ will be discussed below. Only the $O(1)$ 
part of $\Gamma^{\,\gamma}$ affects the quark and gluon densities at order 
$\alpha $, as well as any observable involving hadronic final states like 
$F_2^{\,\gamma}$, leading to the evolution equations~\cite{DWi}
\bea
 \frac{dq_i^{\,\gamma}}{d\ln Q^{2}} &\! =\! & \frac{\alpha}{2\pi} P_{q_i\gamma} 
   \, +\, \frac{\alpha_{s}}{2\pi}\bigg\{ 2\sum_{k=1}^{f} P_{q_{i}q_{k}} \otimes 
   q_{k}^{\,\gamma} \, +\, P_{q_{i}g} \otimes g^{\gamma} \bigg\}     \nn \\
  \frac{dg^{\gamma}}{d\ln Q^{2}} &\! =\! & \frac{\alpha}{2\pi} P_{g \gamma} \:
   \, +\, \frac{\alpha_{s}}{2\pi}\bigg\{ 2\sum_{k=1}^{f} P_{g q_{k}} \:\otimes 
   q_{k}^{\,\gamma} \, +\, P_{gg} \:\otimes g^{\gamma} \bigg\}       \: .
\label{evol3}
\eea
The splitting functions $ P_{ij}(x,\alpha_s) $ are presently known to 
next--to--leading order (NLO) in $\alpha_{s}$, $m = 1$, see refs.~\cite
{FKL,FoPi,GRVt}.

The momentum sum rule (\ref{msr1}) holds order by order in $\alpha$, thus
Eq.~(\ref{evol2}) implies
\beq
\label{msr2}
  \int_0^1 \! dx\, x \Big[ \Sigma_{\rm LO}^{\,\gamma}(x,Q^2) + g^{\gamma}_{\rm 
  LO}(x,Q^2) \Big] = \frac{\alpha}{\pi} \Big( \sum_q e_q^2 \,\ln \frac{Q^2}
  {Q_0^2} + c_1 \Big) \: .
\eeq
The photon's quark and gluon densities are therefore not related by a 
hadron--type sum rule. Instead their momentum fractions rise logarithmically 
with $Q^2$ as long as the lowest--order approximation in $\alpha$ is justified. 
Hence, on the level of Eq.~(\ref{evol3}) alone, an important constraint on the 
parton densities is missing.  That deficit can be removed by inferring $c_1$ 
from elsewhere, as recently done in refs.~\cite{SaS,FrG} by connecting Eq.\ 
(\ref{evol2}) to the cross section $\sigma (e^+e^- \!\ra {\it hadrons})$ via a 
dispersion relation in the photon virtuality. This procedure yields~\cite{SaS}
\beq
\label{msr3}
  \Big(\frac{c_1}{\pi}\Big)_{\rm LO} =  \sum_{V=\rho ,\omega ,\phi } 
  \frac{4\pi}{f_V^2} \:\simeq\:\, 0.55 \:\:\: \mbox{ at } \:\:\: Q_0^2 \:\simeq 
  \: (0.6 \mbox{ GeV})^2 \: .
\eeq
An error of about 20\% can be assigned to this value, arising from the 
uncertainties of $f_{\rho}^2$ (leptonic $\rho$ width vs.\ $\gamma p \ra \rho^0 
p$) and of the scale $Q_0^2$ where the connection of $c_1$ to the vector--meson 
decay constants holds. The numerical results of refs.~\cite{SaS,FrG} agree well 
within this margin.

The general solution of the inhomogeneous evolution equations (\ref{evol3}) 
reads
\beq
\label{sol1}
  \vec{q}^{\,\,\gamma} = \left( \!\! \begin{array}{c} \Sigma^{\,\gamma} 
  \\ g^{\gamma} \end{array} \!\! \right) 
  = \vec{q}_{\,\rm PL}^{\,\,\gamma} + \vec{q}_{\,\rm had}^{\,\,\gamma} \:\: ,
\eeq
where only the flavour singlet part has been indicated. The solution of the 
homogeneous (`hadronic') equation, $\vec{q}_{\,\rm had}^{\,\,\gamma}(x,Q^2)$, 
contains the perturbatively uncalculable boundary conditions $\vec{q}^{\,\,
\gamma} (x,Q_0^2)$. The inhomogeneous (`pointlike') part, on the other hand, 
is completely calculable once $Q_0^2$ has been specified. 

\pagebreak
\noindent 
At next--to--leading order these solutions can be written as~\cite{GRVt,FP}
\beq 
\label{sol2}
  \vec{q}^{\,\,\gamma}_{\,\rm had} = \bigg( \Big[ \frac{\alpha_s}{\alpha_0} 
  \Big]^{\hat{d}} + \frac{\alpha_s}{2\pi} \bigg\{ \hat{U} \otimes \Big[ \frac
  {\alpha_s} {\alpha_0} \Big]^{\hat{d}} - \Big[\frac{\alpha_s}{\alpha_0} \Big]
  ^{\hat{d}} \otimes \hat{U} \bigg\} \bigg) \otimes \vec{q}^{\,\,\gamma}(Q_0^2)
\eeq
and 
\beq
\label{sol3}
  \vec{q}^{\,\,\gamma}_{\,\rm PL} = \bigg\{ \frac{2\pi}{\alpha_s}\! +\! \hat{U}
  \bigg\} \!\otimes\! \bigg\{ 1\! -\!\Big[ \frac{\alpha_s}{\alpha_0} \Big]
  ^{1+\hat{d}} \bigg\} \!\otimes\! \frac{1}{1\! +\!\hat{d}} \otimes \vec{a} + 
  \bigg\{ 1\! -\!\Big[ \frac{\alpha_s}{\alpha_0} \Big]^{\hat{d}} \bigg\} \!
  \otimes\! \frac{1}{\hat{d}} \otimes \vec{b} 
\eeq
with $\alpha_0 = \alpha_s(Q_0^2)$. $\vec{a} $, $\vec{b} $, $\hat{d} $ and 
$\hat{U} $ stand for known combinations of the splitting functions and the QCD 
$\beta$--function. The LO evolution is obtained from Eqs.~(\ref{sol2}) and 
(\ref{sol3}) by putting $\hat{U} = 0$ and $\vec{b} = 0$. A convenient way to 
evaluate these expressions is by transformation to Mellin moments, which 
reduces the convolutions to simple products. The $x$--dependent distributions 
are then calculated by a numerical Mellin inversion of the final result 
(\ref{sol1}).
\section{Boundary conditions and factorization schemes}
The structure function $F_{2}^{\, \gamma} $ is, at first order in the 
electromagnetic coupling $\alpha$,
\beq
\label{f2ph}
  F_{2}^{\,\gamma} \: = \!\sum_{q=u,d,s} \! 2x\, e_{q}^{2} \,\Big\{ q^{\gamma}
  + \frac{\alpha_{s}}{2\pi} \big( C_{2,q} \otimes q^{\gamma} + C_{2,g} \otimes 
  g^{\gamma} \big) + \frac{\alpha}{2\pi} e_{q}^{2}\, C_{2,\gamma} \Big\} \: .
\eeq
Only the contribution of the light flavours has been written out here. The 
reader is referred to refs.~\cite{LRSN,LR} for the heavy quark part $F_{2,\, h}
^{\,\gamma}$. At LO in $\alpha_s$, just the first term in Eq.~(\ref{f2ph}) is 
taken into account since $q^{\gamma}\sim 1/\alpha_{s\,}$, see Eq.~(\ref{sol3}). 
At NLO the usual hadronic one--loop coefficient functions $C_{2,q}(x)$ and 
$C_{2,g}(x)$ enter, together with the direct--photon contribution 
$C_{2,\gamma}$ given by~\cite{BB} 
\beq
\label{c2ph}
  C_{\, 2,\gamma}^{\,\overline{\rm MS}}(x)  = 3 \Big( \big[ x^2 + (1-x^2) \big] 
  \ln \frac{1-x}{x} - 1 + 8x (1-x) \Big) \: . 
\eeq
This term causes difficulties in this standard factorization scheme, as it 
leads to a large LO/NLO difference for the inhomogeneous part $F_{2,\,\rm PL}^
{\, \gamma}$. In particular it is strongly negative at large $x$, see Fig.~1. 
Thus $F_{2,\,\rm PL}^{\,\gamma}$ turns positive over the full $x$-range, for
$Q_0^2 = 1 \mbox{ GeV}^2$, only at $Q^2 \simeq 20 \mbox{ GeV}^2$. This 
unphysical behaviour has to be overcome in the complete $F_{2}^{\,\gamma}$ by 
the \MSb\ initial distributions, which are therefore forced to be very 
different from their LO counterparts. 
\begin{figure}[tb]
\centerline{\mbox{\epsfig{file=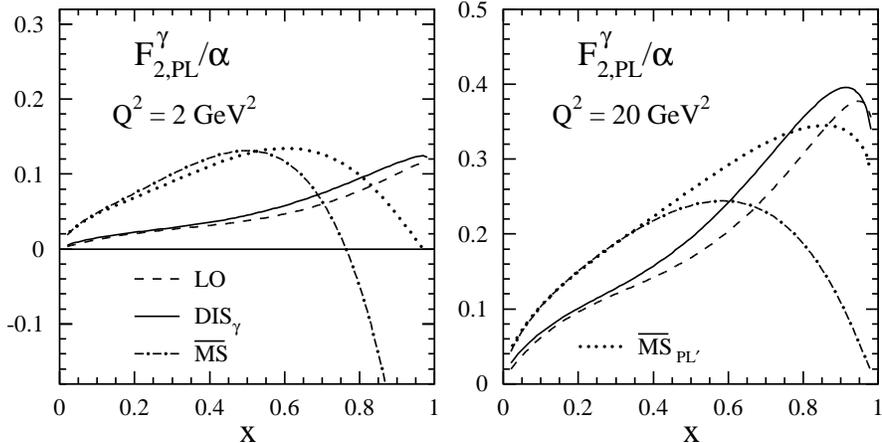,width=11.8cm}}}
\vspace*{-3mm}
\caption{The pointlike structure function $F_{2,\,\rm PL}^{\,\gamma}$ in LO and
in NLO for the \MSb\ and \DISg\ schemes. $Q_0^2 = 1 \mbox{ GeV}^{2}$, three 
active flavours and $\Lambda_{\rm LO} = \Lambda_{\rm NLO}= 0.2 \mbox{ GeV}$ 
have been used.}
\vspace*{-3mm}
\end{figure}

These problems are circumvented by adopting the \DISg\ scheme introduced in 
refs.~\cite{GRVt,GRVg}. Here $C_{2,\gamma} $ is absorbed into the quark 
distributions according to 
\beq
\label{disg}
  q^{\,\gamma}_{\,{\rm DIS}_{\gamma}} = q^{\,\gamma}_{\,\overline{\rm MS}} 
  + \frac{\alpha}{2\pi} e^{2}_{q}\, C_{\, 2,\gamma}^{\,\overline{\rm MS}} 
  \: , \:\:\:\:  C_{\, 2,\gamma}^{\,{\rm DIS}_{\gamma}} = 0  \: .
\eeq
The coefficient functions $C_{2,q}$ and $C_{2,g}$ in Eq.~(\ref{f2ph}), as well 
as the definition of the gluon density remain unchanged, in contrast to the 
hadronic DIS scheme~\cite{AEM}. Therefore $F_{2}^{\,\gamma}$ assumes the usual 
hadronic \MSb\ form without a direct term in \DISg , resulting in a good LO/NLO 
stability of $F_{2,\,\rm PL}^{\,\gamma}$ as illustrated in Fig.~1. Consequently 
physically motivated boundary conditions for the quark and gluon densities can 
be employed in this scheme also beyond leading order.
An additional advantage of the \DISg\ scheme is that the leading \MSb\ terms 
for $ x \ra 0 $ cancel in the transformed NLO photon--parton splitting 
functions~\cite{GRVf}
\beq
\label{pdis}
  P_{q \gamma}^{\,(1)} \sim \bigg\{ \! \begin{array}{r} \ln ^{2} x + \ldots 
  \\ 2 \ln x + \ldots \end{array} \: , \:\:\:\:
  P_{g \gamma}^{\,(1)} \sim \bigg\{ \! \begin{array}{rl} 1/x + \ldots 
   & $ \MSb $ \\ -3 \ln x + \ldots & $ \DISg $ 
  \end{array} \: .
\eeq
In fact, this cancellation of the leading small--$x$ term of $P_{g \gamma}^{\,
\overline{\rm MS}}$ does not only take place at NLO, but persists to all orders 
in $\alpha_s$~\cite{BV}.

\begin{figure}[tb]
\centerline{\mbox{\epsfig{file=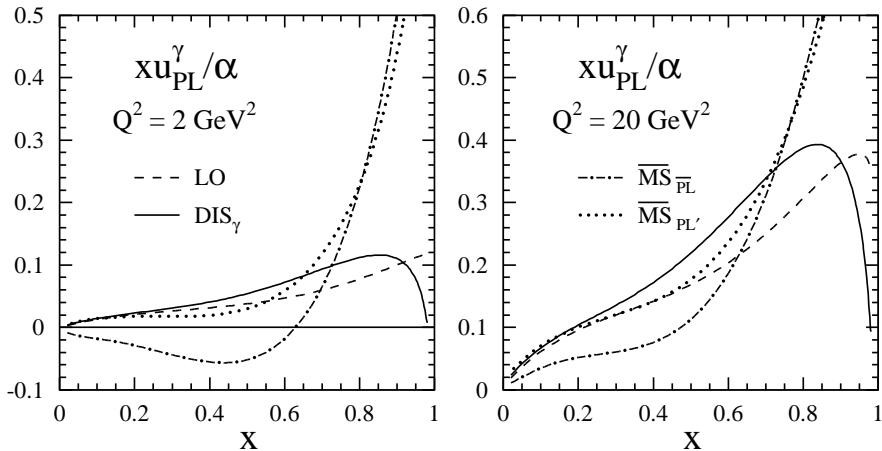,width=11.8cm}}}
\vspace*{-3mm}
\caption{The pointlike up--quark density $ u_{\,\rm PL}^{\gamma} $ in LO and 
in NLO for the \DISg\ scheme, compared to the physically equivalent \MSb\ 
distribution ($\overline{\mbox{PL}}$) with the input (\ref{msbg}). The 
parameters $Q_{0}^{2}$, $f$ and $\Lambda$ are as in Fig.~1. Also shown is the 
result for the PL$'$ boundary condition (\ref{iafg}).}
\vspace*{-3mm}
\end{figure}
An equivalent \MSb\ formulation of the above solution to the $C_{2,\gamma}$
problem has been pursued in refs~\cite{GS92,GS96}. It can be written as a 
modification of the pointlike part ($\overline {\mbox{PL}}$) in Eq.~(\ref
{sol1}) by an additional `technical' NLO input density,
\beq
\label{msbg}
  q_{\,\overline{\rm PL}}^{\gamma}(x, Q_{0}^{2}) = - \frac{\alpha}{2\pi} 
  e^{2}_{q} \, C_{2,\gamma}^{\,\overline{\rm MS}}(x) \: , \:\:\:\:
  g_{\,\overline{\rm PL}}^{\gamma}(x, Q_{0}^{2}) = 0 \: .
\eeq
This leads to $ F_{2,\, \overline{\rm PL}}^{\,\gamma}(x,Q_0^2) = 0$ and thus 
allows for similar `physical' initial distributions on top of Eq.~(\ref{msbg}). 
The resulting quark distributions, however, exhibit a rather unphysical shape. 
As displayed in Fig.~2, they are suppressed (strongly enhanced) at medium 
(large) values of $x$, respectively, with respect to the pointlike LO and 
\DISg\ results. In a fully consistent NLO calculation, this \MSb\ procedure is 
nevertheless strictly equivalent to the \DISg\ treatment. On the other hand, as 
soon as not all terms beyond NLO are carefully omitted, the \MSb\ treatment
turns out to be unstable at large $x$, see refs.~\cite{GRVa,AV94}.

Due to the non--universality of the coefficient function $C_{2,\gamma}$, a 
special role is assigned to $F_2^{\,\gamma}$ in the redefinitions (\ref
{disg},\ref{msbg}) of the quark densities, similar to the hadronic DIS scheme. 
An alternative process--independent approach was worked out in ref.~\cite{AFG}. 
A universal technical \MSb\ input has been inferred from a detailed analysis 
of the Feynman diagrams for $\gamma^{\ast} \gamma \ra \gamma^{\ast} \gamma $,
which leads to
\beq
\label{iafg}
  q_{\,{\rm PL}^{\prime}}^{\,\gamma}(x,Q_{0}^{2}) = - \frac{\alpha}{2\pi} 
  e^{2}_{q}\, C^{\,\prime}_{\gamma}(x) \: , \:\:\:\: g_{\,{\rm PL}^{\prime}}
  ^{\,\gamma}(x,Q_{0}^{2}) = 0 
\eeq
with
\beq
\label{c2pr}
  C^{\,\prime}_{\gamma}(x) = 3 \Big( \big[ x^2 + (1-x^2) \big] \ln (1-x) 
  + 2x (1-x) \Big) \: . 
\eeq
The resulting modified pointlike structure function $ F_{2,\,{\rm PL}^{\prime}}
^{\,\gamma} $, also shown in Fig.~1, remains negative at large $x$ due to the 
uncompensated $-1$ in Eq.~(\ref{c2ph}) only close to the reference scale 
$Q_0^2$. At medium to small $x$, $ F_{2,\,{\rm PL}^{\prime}}^{\,\gamma}$ is 
similar to the pointlike \MSb\ results, i.e., larger than its LO and \DISg\ = 
{\small $\overline {\mbox{PL}}$} counterparts. The corresponding up--quark 
distributions are also illustrated in Fig.~2.
\section{Parametrizations of photonic parton distributions}
\vspace*{-0.5mm}
In order to specify the photon's parton densities, the perturbatively 
uncalculable initial distributions, $q_i^{\gamma}(x,Q_0^2)$ and $g^{\gamma}(x,
Q_0^2)$, have to be fixed at some scale $Q_0^2$. Only one combination of quark 
densities (dominated by $u^{\gamma}$) is presently well constrained, however, 
by $F_{2}^{\,\gamma}$ data at $0.01 \,\lsim\, x\, \lsim\, 0.8$ from PETRA 
\cite{PLUT,JADE,TASS}, PEP~\cite{TPC}, TRISTAN~\cite{AMY,TOPA}, and LEP~\cite
{OPAL,DELP,ALEP}. The complete present data, including new results presented at 
this conference, is shown in Fig.~3 together with the NLO parametrizations of
refs.~\cite{GRVg,AFG}. The gluon distribution is not tightly constricted 
either: there is sound evidence for $g^{\gamma} \ne 0 $, and a very large and 
hard $ g^{\gamma}$ has been ruled out by jet production results~\cite
{TOPj,AMYj,H1j}. 
\begin{figure}[tp]
\centerline{\mbox{\epsfig{file=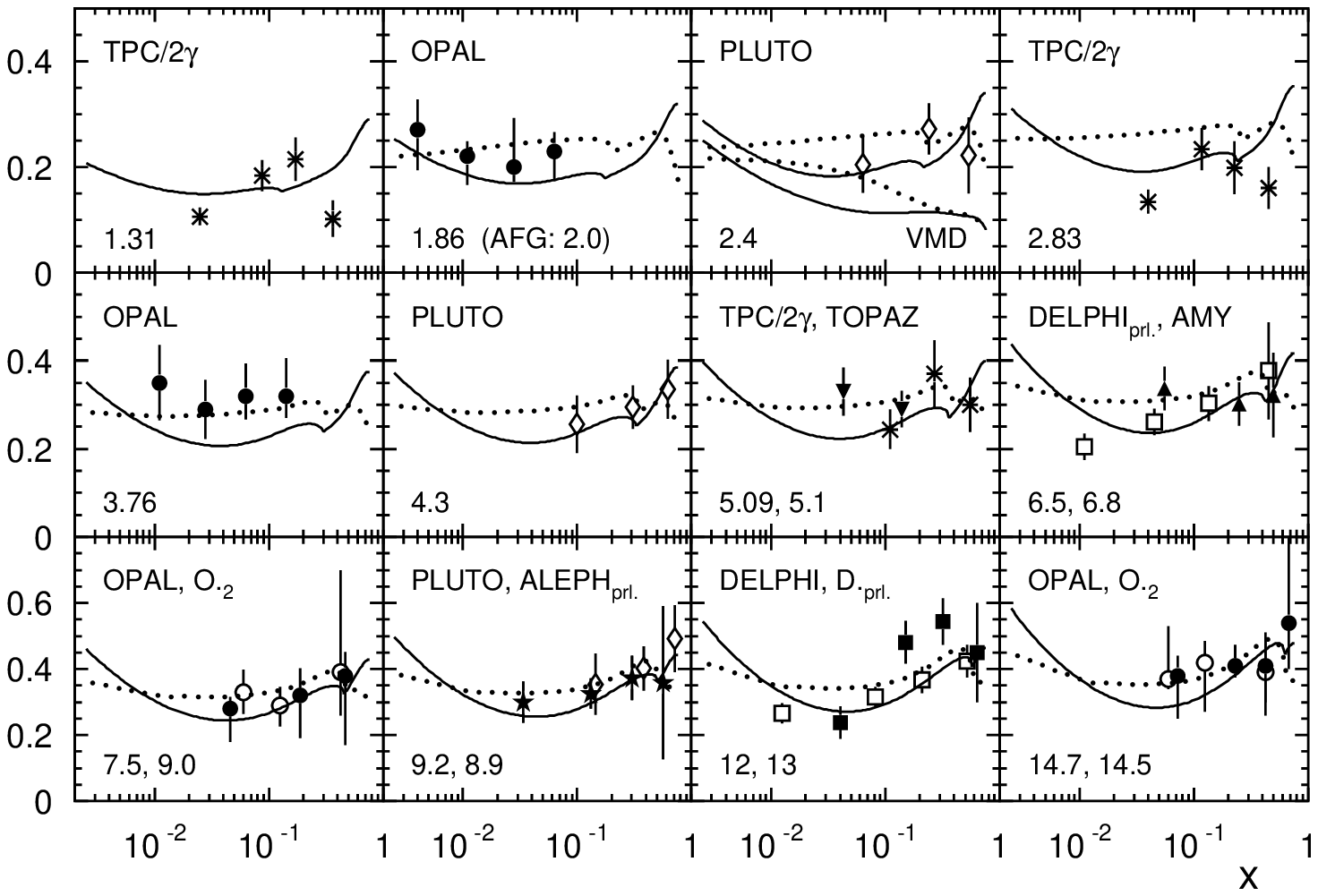,width=11.8cm}}}
\centerline{\mbox{\epsfig{file=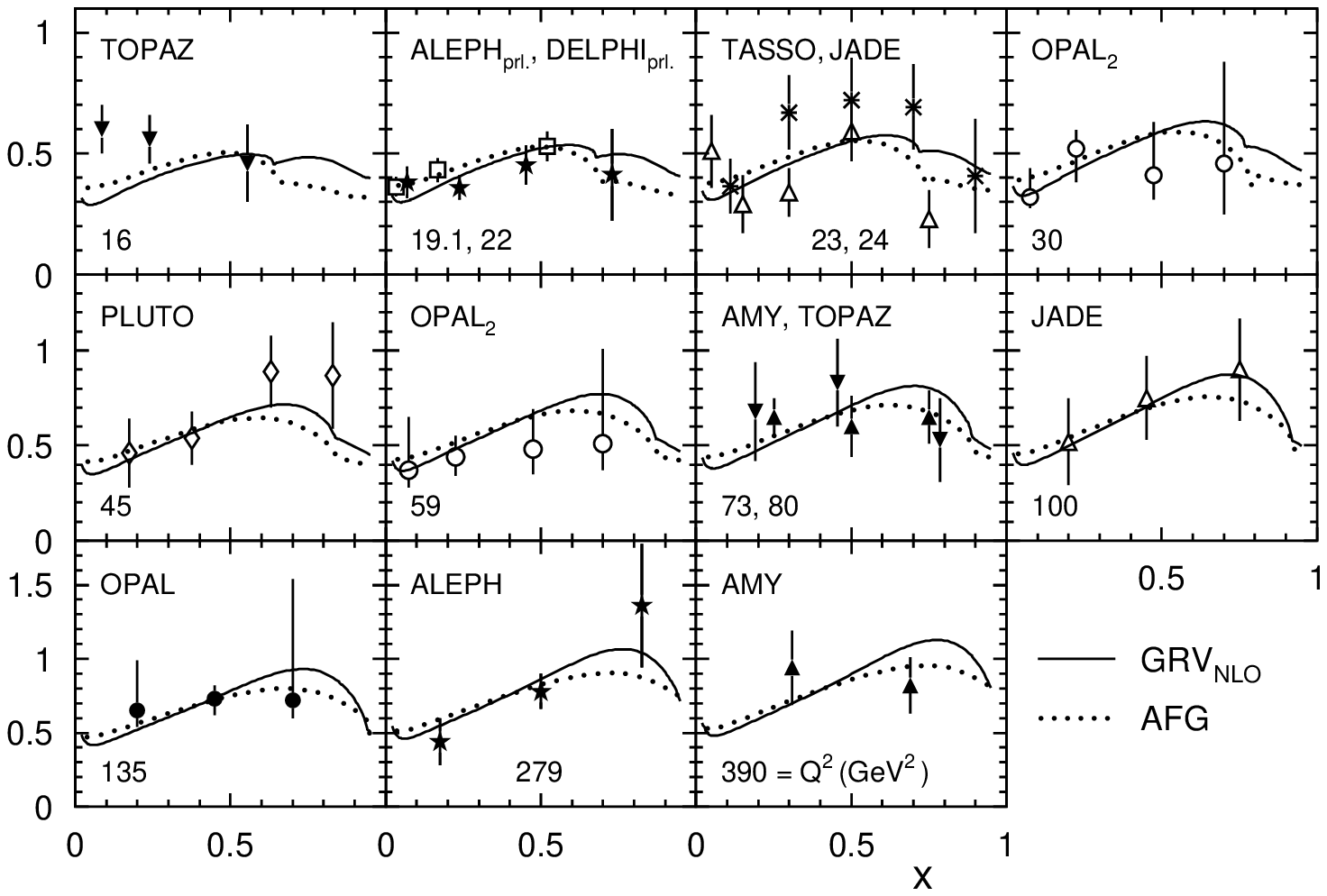,width=11.8cm}}}
\vspace*{-3.5mm}
\caption{The presently available $F_{2}^{\gamma}$ data compared to NLO
parametrizations of refs.~\protect\cite{GRVg,AFG}. The hadron--like VMD 
components of the latter are separately displayed at $Q^2 = 2.4\mbox{ GeV}^2$.}
\end{figure}

Due to these limitations, current parametrizations invoke theoretical 
estimates and model assumptions, in particular from vector meson dominance 
(VMD). For safely high reference scales, $Q_0^2 \,\gsim\, 2 \mbox{ GeV}^2$,
however, purely had\-ron--like initial distributions are known to be 
insufficient. An additional hard quark component has to be supplemented there 
in order to meet the $F_2^{\,\gamma}$ data at larger $Q^2$. 
In view of this situation two approaches have been used. First one can keep 
$ Q_0 \geq 1 \mbox{ GeV} $, fit the quark densities to $ F_2^{\,\gamma}$ data, 
and estimate the gluon input. This method has been adopted in refs.~\cite
{DG,LAC} and, more recently, in refs.~\cite{SaS,GS96,WHIT,AGL}. The second 
option is to retain a pure VMD ansatz, 
\beq
\vspace*{-1mm}
\label{vmd}
  (q_i^{\,\gamma}, g^{\gamma})(x,Q_0^2) = \frac{4\pi \alpha}{f_{\rho}^2} \,
  (q_i^{\,\rho}, g^{\,\rho}) (x,Q_0^2) + \ldots \: ,
\eeq
together with assumptions on the experimentally unknown $\rho $ distributions,
and to start the evolution at a very low scale $Q_0 \simeq 0.5 \ldots 0.7 
\mbox{ GeV}$~\cite{SaS,GRVg,AFG,Aurg}. Note that this boundary condition 
complies with the momentum sum rule (\ref{msr3}) if the $\omega $ and $\phi $
contributions are appropriately added.

In the following, the three available NLO parametrizations~\cite{GRVg,GS96,AFG} 
are briefly compared, together with the recent LO sets of ref.~\cite{SaS}. For 
all these distributions $\Lambda_{{\rm LO},\,\overline{\rm MS}} = 200 \mbox{ 
MeV}$ have been employed at $f=4$.

\begin{figure}[t]
\centerline{\mbox{\epsfig{file=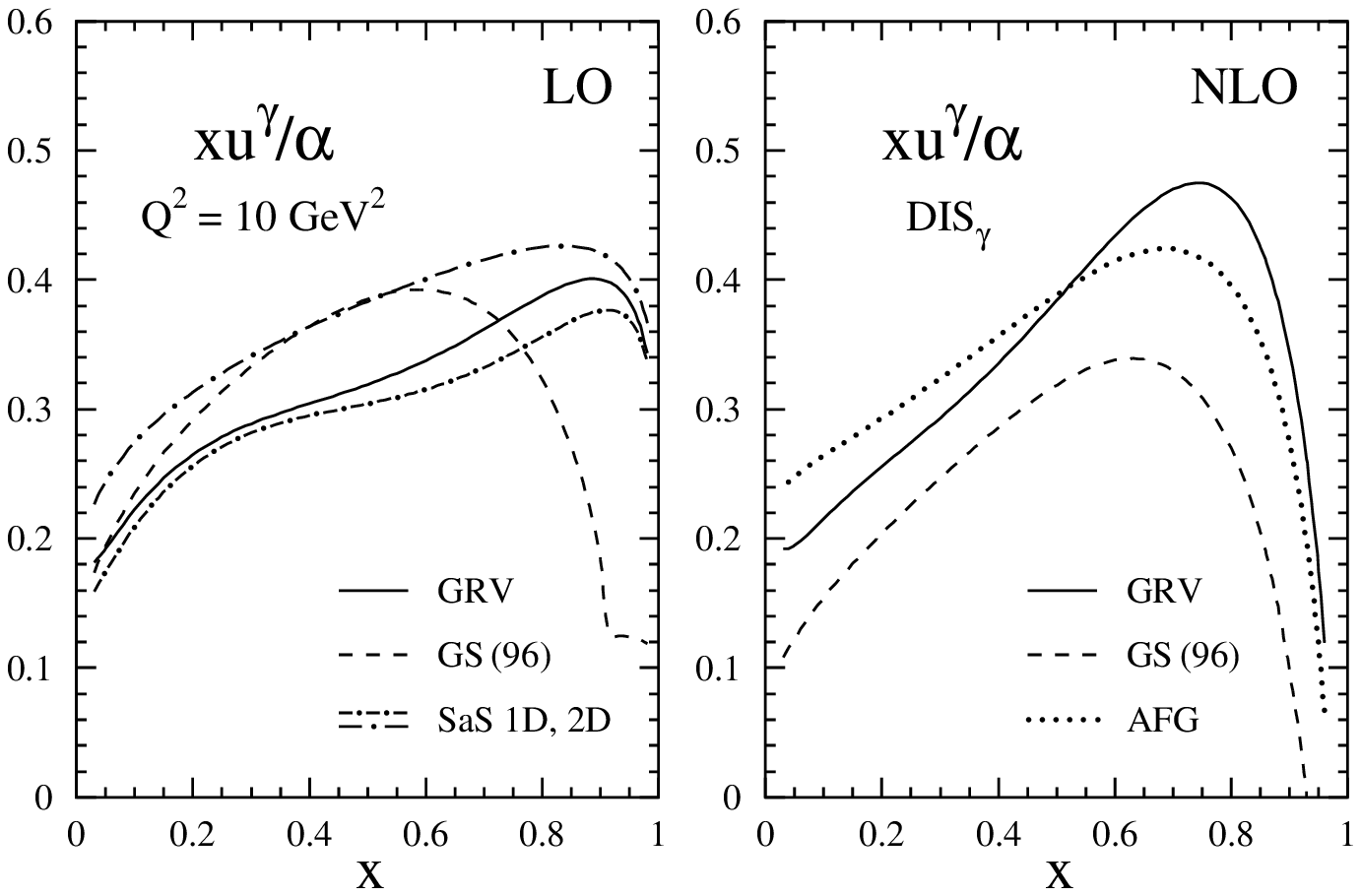,width=11.2cm}}}
\vspace*{-3mm}
\caption{Parametrizations of the up--quark distribution at LO~\protect\cite 
{SaS,GRVg,GS96} and NLO~\protect\cite{GRVg,GS96,AFG}. NLO results in \MSb\ have
been transformed to the \DISg\ scheme according to Eq.~(\ref{disg}).}
\vspace*{-3mm}
\end{figure}
The resulting $u$--quark densities are displayed in Fig.~4. Considering the LO 
results first, one notices that the parametrizations form two groups in the
well--measured intermediate $x$-range, $ 0.2 \,\lsim\, x \lsim 0.7 $. The lower
one consists of the two low--$Q_0$ sets, GRV~\cite{GRVg} and SAS$\,$1D~\cite
{SaS}, which start the evolution at $Q_0^2 = 0.25 \mbox{ GeV}^2$ and $0.36 
\mbox{ GeV}^2$, respectively. The reference scales for the higher SAS$\,$2D
\footnote{The additional SaS$\,$1M and SaS$\,$2M sets in ref.~\cite{SaS} are 
theoretically inconsistent, as the leading--order evolution is combined with 
the scheme--dependent coefficient function $C_{2, \gamma}$.}, and GS$\,$(96)
\cite{GS96} distributions read $Q_0^2 = 4 \mbox{ GeV}^2 $ and $3 \mbox{ GeV}
^2$. This difference has been driven at least partly by first LEP data~\cite
{DELP,OPAo}, see ref.~\cite{SSV}, which were considerably higher than previous 
results around $x = 0.2$. A more consistent picture is now emerging from the 
new LEP data in this range.

The second striking feature in Fig.~4 is the large--$x$ behaviour of the GS
parametrization. Unlike SaS$\,$2D, where a simple hard term $\propto x$ is 
employed, GS choose the massive Born expressions for $\gamma^{\ast} \gamma 
\ra q \bar{q}$ at $Q_0^2$ on top of the hadronic VMD input. All power--law 
contributions $O(\, [m_q^2/Q^2]^{n\, })$, $n \geq 1$, are retained, resulting 
in a threshold at $x \simeq 0.9$ for typical constituent quark masses. Such a 
procedure, however, may be considered as inadequate for the construction of 
leading--twist parton densities.

Let us now turn to the NLO distributions. The results of GRV~\cite{GRVg} and
AFG~\cite{AFG} are both based on the VMD ansatz (\ref{vmd}), imposed at $Q_0^2 
= 0.3 \mbox{ GeV}^2$ and $0.5 \mbox{ GeV}^2$, respectively. The differences 
between these two parametrizations at $x \gsim 0.1 $ can be understood in terms
of the non--hadronic NLO boundary conditions discussed in Sec.~3, cf.~Fig.~1.
At lower $x$, the deviations are dominated by the differing assumptions~\cite
{ABFK,GRVp} on the experimentally virtually unconstrained pion sea -- both 
groups estimate the unknown $\rho$ distributions by their pionic counterparts. 
The third NLO set, GS$\,$(96), is technically flawed: it should at $Q_0^2$, by 
construction, lead to the same $F_2^{\,\gamma}$ results as the LO fit.
However, $u^{\gamma}$ turns out to be sizeably too small over the full 
$x$--range. Hence this parametrization is unfortunately not usable in its 
present form~\footnote{This discussion also applies to the previous NLO
parametrization~\cite{GS92} of the same group.}. 

\begin{figure}[t]
\centerline{\mbox{\epsfig{file=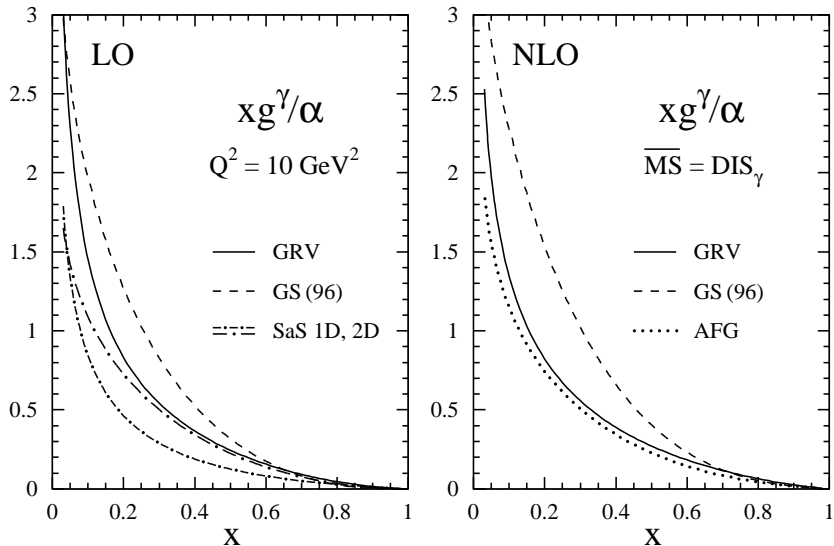,width=11.2cm}}}
\vspace*{-4mm}
\caption{Parametrizations of the photon's gluon density at LO~\protect\cite
{SaS,GRVg,GS96} and NLO~\protect\cite{GRVg,GS96,AFG}.}
\vspace*{-4mm}
\end{figure}
Before considering the gluon density, it is appropriate to comment on the
quark--flavour decomposition and the momentum sum rule. Both SaS$\,$1D and AFG 
perform a coherent addition of the three light vector mesons at their
respective input scales, with slightly differing assumptions on SU(3) breaking 
and the value of $f_{\rho}$. That leads to a suppression of the $d$--valence 
density by a factor of four with respect to the $\rho$--meson's $u$--valence 
component. This approach is able to describe the $F_2^{\,\gamma}$ data without
any further adjustment, hence the momentum sum rule (\ref{msr3}) is met in both 
cases. On the other hand, GRV use just a $\rho$ distribution, with a prefactor 
adjusted to the data. Although a factor of 1.6 perfectly mimics the $F_2$ of 
the (SU(3) symmetric) coherent superposition, too much momentum is spent due
to the $u_v = d_v$ symmetry. Thus Eq.~(\ref{msr2}) is violated, e.g., by about
+40\% in LO at $Q^2 = 4 \mbox{ GeV}^2$. Finally the high--$Q_0$ fits, SaS$\,$2D
and GS, do not impose the momentum sum rule at all. 

The gluon distributions of these parametrizations are finally presented in 
Fig.~5. The pion distributions of AFG~\cite{ABFK} and GRV~\cite{GRVp} both 
describe the direct--photon production data in $\pi p$ collisions~\cite{WA70}, 
that is why these photonic gluons are so similar except at very $x$. For the GS
parametrization, the gluon densities have been constrained by a LO comparison
to TRISTAN jet production data~\cite{TOPj,AMYj}, which seem to prefer a 
relatively large gluon distributions. The shapes of the SaS gluon densities
are fixed by theoretical estimates, no direct or indirect experimental
constraint has been imposed here.
\section{Photon structure at small {\boldmath $x$}}
\vspace*{-0.5mm}
The region of very small parton momenta, $10^{-5} \lsim x \lsim 10^{-2}$, has 
attracted considerable interest in the proton case since the advent of HERA. 
The quark and gluon distributions show a marked rise at small~$x$~\cite
{H1pr,ZEpr}, in good agreement with perturbative predictions for a low input
scale $Q_0 \simeq 600 \mbox{ MeV}$~\cite{GRVn}. The corresponding NLO evolution 
of the photon structure is shown in Fig.~6.
\begin{figure}[htb]
\vspace*{-2mm}
\centerline{\mbox{\epsfig{file=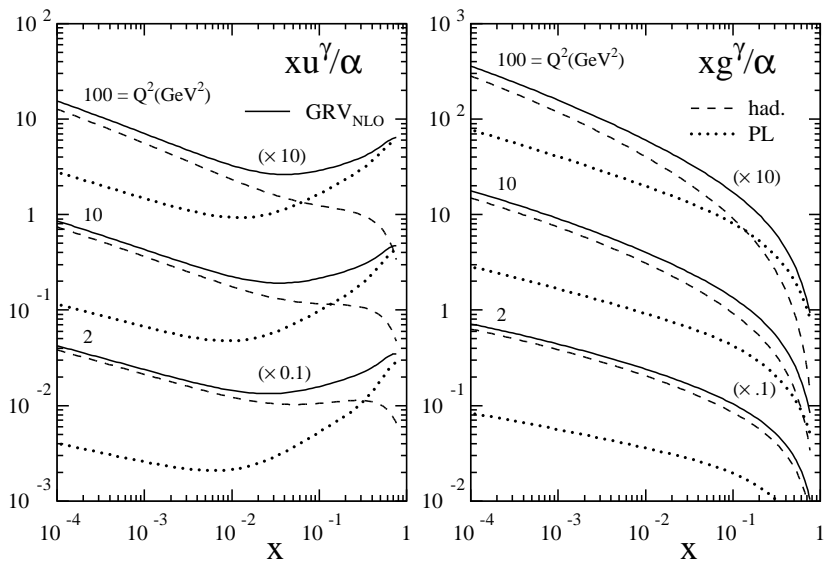,width=11.0cm}}}
\vspace*{-4mm}
\caption{The NLO small-$x$ evolution of the photon's quark and gluon 
distributions as predicted in ref~\protect\cite{GRVg}. The hadronic (VMD) and
pointlike contributions are shown separately.}
\vspace*{-1mm}
\end{figure}
 
The parton distributions of the photon 
behave very differently in the limits of large and small $x$. In the former 
case, the perturbative part (\ref{sol3}) dominates, especially for the quark 
distributions. On the other hand, this calculable contribution amounts at most 
to about 20\% at very small-$x$, at scales accessible in the foreseeable 
future, in LO as well as in NLO. One may therefore expect, by VMD arguments for 
the hadronic component (\ref{sol2}), a very similar rise as observed in the 
proton case. It will be very interesting to see whether this expectation is 
borne out by future $F_2^{\,\gamma}$ measurements.

Let me finally mention that there is an even more intriguing, if exotic, 
possibility here: the calculable pointlike contribution could be drastically 
enhanced by large logarithmic small-$x$ terms in the perturbation series~\cite
{BV}. If this component could be projected out, for example, by final--state 
observables, it would provide a rather unique small-$x$ QCD laboratory.
\section*{Acknowledgement}
This work has been supported by the German Federal Ministry for Research and 
Technology (BMBF) under contract No.\ 05 7WZ91P (0).
\end{document}